# Gellan-Based Hydrogels and Microgels for culturage heritage: a rheological perspective

Silvia Franco[a,b,*], Leonardo Severini[a,b], Elena Buratti[c], Letizia Tavagnacco[a,b], Simona Sennato[a,b], Mauro Missori[a,b], Barbara Ruzicka[a,b], Claudia Mazzuca[d], Emanuela Zaccarelli[a,b,*], Roberta Angelini[a,b,*]

[a]Institute for Complex Systems, National Research Council, Piazzale Aldo Moro, 5, Rome, 00185, Italy
[b]Physics Department, Sapienza University of Rome, Piazzale Aldo Moro 5, Rome, 00185, Italy
[c]Department of Chemical, Pharmaceutical and Agricultural Sciences, University of Ferrara, Via L. Borsari 46, Ferrara, 44121, Italy
[d]Department of Chemical Science and Technologies, University of Rome Tor Vergata, Via Della Ricerca Scientifica 1, Roma, 00133, Italy

## Abstract

Gellan gum has gained significant attention due to its versatility in multiple applications in the form of hydrogels and microgels. A thorough understanding of the rheological behaviour of these systems is crucial both for fundamental research and to optimize the manufacturing needs. To this aim, here we extensively characterize the rheological behaviour of gellan based hydrogels and microgels recently used for efficient paper cleaning for restoration interventions. In particular, we study their viscoelastic properties, also during hydrogel and microgel formation, assessing the role of temperature, gellan concentration, and, importantly, the presence of different cations, which plays a crucial role in the gelation process. We find the interesting result that, inthe conditions where they are efficient for cleaning, gellan hydrogels exhibita double yielding behavior. In addition, we provide a detailed description of gellan microgels preparation, ensuring high control and reproducibility of the samples. Altogether our study sheds light on the mechanical stability, net-work structure, and overall functionality of the gellan-based gels, providing valuable insights into optimizing conditions for desired applications in paper cleaning.

---

*Corresponding author
Email addresses: silvia.franco@cnr.it (Silvia Franco),
emanuela.zaccarelli@cnr.it (Emanuela Zaccarelli), roberta.angelini@cnr.it
(Roberta Angelini)

## 1. Introduction

Gellan gum (GG) is a polysaccharide derived from microbial fermentation of sugars by the bacterium Sphingomonas elodea and commonly employed in manufacturing sectors such as food, pharmaceutical, and cosmetic industries, due to its gelling, stabilizing, and thickening properties [1, 2, 3] . Its repeating unit consists of a tetramer of one (1,3)-*β*-D-glucose, one (1,4)-*β*-D-glucose, one (1,3)-*β*-D-glucuronic acid, and one (1,4)-*α*-L-rhamnose [1] . The ability of GG to form stable physical gels, its compatibility with a wide range of ingredients, the versatility in modifying texture and consistency as well as its easiness in chemical modification have contributed to its widespread use in many fields. In particular, for food processing and production, it serves as a gelling agent, stabilizer, and thickener, enhancing the texture and stability of products like jams, jellies, and dairy alternatives [4, 5, 6] . In pharma- ceutics and biomedicine, GG is utilized for controlled drug delivery [7, 8, 9] , as a delivery system for superficial cutaneous administration [10] , in tissue engineering [11] , and as a bioink in 3D printing to create cell-laden con- structs [12] . Its ability to form strong, clear gels makes it valuable also in cosmetic formulations for products such as lotions and creams [13] . Addi- tionally, GG is used in the agricultural sector as fertilizer-release agents [14] , for seed coating and as a component in plant tissue culture media.
It should be pointed out that commercially Gellan gum is available in differ- ent forms. Gellan gum in its native form ("high acyl" one) has two acyl moi- eties, L-glyceryl and acetyl groups, linked to the (1,3)-*β*-D-glucose residue. In the other form the acyl groups are partially or totally removed by alka- line hydrolysis. The acylation degree affects the rheological and mechanical properties of the resulting physical hydrogel: "high acyl" GG leads to obtain soft, elastic, and opaque hydrogels, while deacylation allows to obtain rigid and compact hydrogels, characterized by a high degree of viscoelasticity and a remarkable transparency [4, 15, 16, 17]. It is well-known from the litera-ture that deacylated GG sol-gel transition is an exothermic process [18]: GG chains, dispersed in aqueous solution and subsequently heated ($\geq 90°$), assume a random coil conformation; subsequently, in the cooling process, the chains undergo a transition from a more disordered state to a more ordered one, through the formation of single or double helices by spontaneous self-assembly processes (coil-helix transition) [19, 20] . Upon further cooling, there is an association between the helical structures, leading to the sol-gel transition. The rheological properties of the so-formed hydrogels are strongly dependent on several factors like polysaccharide concentration, working temperature, charge and type of cations added in the system [16, 18, 21] . In detail, metal cations are able to shield electrostatic repulsion forces between polysaccharide helices (due to the presence of glucuronic acid's carboxylated groups) in the so called "junction zones", favoring chain aggregation [18] . Notwithstanding the widespread use of GG in very different fields, rheological measurements on gellan gum-based systems are present in the literature only for systems with relatively low GG concentration (up to 1%), either in the presence or in the absence of salts [22, 23, 24] . This is a critical lack of data because the new frontiers opened by various applications of this polysaccharide and the potentiality offered by the chemical modifications of its residues require a deep knowledge of the rheological properties of gellan gum based gels at higher concentrations in order to optimize their use. In addition, recent works have put forward the employment of GG microgels [25, 26]·, whose detailed rheological properties have not been investigated so far. In particular, one promising field of application of GG gels is given by cultural heritage. Indeed, in the last few years, GG based hydrogels have emerged as efficient wet cleaning treatments for paper artworks [27] . Wet cleaning is a very delicate operation, and to be used for this purpose, hydro- gels must fulfill several strict conditions from the mechanical point of view:

i) they have to be easily handled, applied and removed on paper sheets to be cleaned without breaks and leaving residues; ii) they have to be able to stand quite significant pressures without breaking. This is because a weight is usually applied on the hydrogel to ensure a close contact between the gel and the artwork to be cleaned. Typically, for a round portion of a gel with 3 cm of radius, a weight of about 150 g may be used; this means that the pressure on the gel is about of 520 Pa [28] . At the same time, they have to be soft enough, so that the solvent can effectively exert its cleaning action without wetting the sample too much and then be removed together with the gel itself. Therefore, a deep knowledge of the rheological properties of gels to be used for cleaning is therefore of fundamental importance to avoid unwanted damages on artworks. Despite this need, only a few characterizations of the rheological properties of GG based gels usable for cultural heritage have been provided [29, 30] . In this context, for example, it is also important to investigate the gelation mechanism for different polymer and salt concentrations. Indeed, at low GG concentration, gelation involves first the formation of GG single helices (or disordered structures), and then coil-coil aggregation [31, 19, 20, 32] leading to extended aggregates. This mechanism could lead, according to Robinson et al. [33] to weak gels. At the same time, we have recently demonstrated by molecular dynamics simulations that, at higher polysaccharide concentration, the first gelation step is the formation of double helices and then superstructures occur through aggregation of double helices. Both these processes are strongly mediated by the presence and nature of cations involved [34] . It is important to complement these studies with a thorough rheological characterization, also to put GG hydrogels in relation to different types of materials. For example, in the presence of a two-step aggregation mechanism, it is legitimate to expect the presence of characteristic rheological features such as double yielding [35]. This has often been observed in the context of colloidal gels [36, 37], but only a few hydrogel evidences have been provided [38] . The present work will also address such important issue for the different hydrogels prepared for paper cleaning.

In addition, we recently put forward the use of GG gel microgels as cleaning tools in the cultural heritage field [25] , as they ensure a very fast clean- ing with efficacy close or higher than the corresponding GG hydrogels. To this aim, we have developed a reliable and reproducible protocol for making microgels, also effective in paper cleaning, based on the use of a rheometer-controlled procedure, that we detail in this work. In addition, we recently tested the efficacy of methacrylated gellan gum hydrogels and microgels. Due to the high amount of hydroxyl moieties, GG can be easily chemically derivatized for obtaining polymers with peculiar features. In this case methacrylation allows the increase of the hydrophobicity of polysaccharide chains; as a consequence the use methacrylated gellan gum hydrogels or microgels as wet cleaning materials allows the simultaneous removal of both hydrophilic (i.e. acids deriving from cellulose degradation) and hydrophobic (oil stains, adhesives) materials from paper surfaces. For all these cases, an extensive rheological characterization is lacking. This is the aim of the present work which reports a systematic and methodological investigation of the viscoelas-tic properties of GG hydrogels and microgels employed in artwork restoration [30, 25] . Our extensive characterization is organized into four main results sections. First, we start by discussing GG hydrogels at high polysac- charide concentration, content and type (monovalent and divalent) of added salts. Then we move on to the case of methacrylated GG hydrogels, where we vary GG and salt content in order to optimize the rheological performance of the gel to be suitable for paper cleaning treatments. Next, we turn to microgels. After discussing our newly established, reproducible preparation protocol, we then characterize their viscoelastic properties as a function of concentration, first for pure GG and then for methacrylated one. All discussed cases refer to conditions where we already tested the ability of the hydrogels and microgels, either methacrylated or not, to efficiently clean paper artworks. After presenting all our results, we then draw conclusions and discuss perspectives for the future.

## 2. Materials and Experimental Methods

### 2.1. Materials

Low acyl Gellan gum (GG) powder Kelcogel$^{TM}$ , was from CP Kelco (San Diego, California) with molecular weight $2.5 \cdot 10^5$ Daltons and dry substance content (DS content) 4-5%. Glycidyl methacrylate, sodium chloride, calcium acetate hydrate, sodium oxide were from Merk (Merk KGaA, Darmstadt, Germany). Reagents used in this work were of analytical grade and used without further purification. Ultrapure water (resistivity: 18.2 M$\Omega$/cm at 25 °C), obtained with Arium® pro Ultrapure water purification Systems, Sartorius Stedim, was used for the solution preparation. GG methacrylation occurred as reported previously [30] .

### 2.2. Hydrogel Preparation

GG and GGMA hydrogels were prepared following the same protocol as ref. [29, 39, 40, 7] Pure or methacrylate GG was dissolved in ultrapure water under stirring at room temperature. The amount of GG powder was weighed according to the final desired concentration. Since gelation mechanism is influenced both by temperature and by the presence of cations in water that makes the gel hard and brittle with a more ordered "crystalline-like" structure [40] , a proper amount of sodium chloride (NaCl) or calcium acetate (Ca(CH$_3$COO)$_2$, Ac$_2$Ca) solution (1 M) was added to the solution to obtain the established salt concentration, C$_s$. The resulting mixture was heated rapidly to the boiling point becoming transparent, then the homogeneous solution was poured into the Petri dishes and left to cool at room temperature for at least an hour. In this work pure GG hydrogels were prepared at different weight concentrations ($C_w$ = 0.5%, 1.0%, 1.5%, 2.0%, 2.5%, 3.0%, 4.0%, 5.0%) while GG hydrogels with the addition of NaCl ($C_{NaCl}$ = 0.5, 2.5, 5.0, 27.0 mM) or of $Ac_2Ca$ ($C_{Ac_2Ca}$ = 0.25, 1.25, 2.5, 5.0 mM) were also prepared at fixed GG concentration $C_w$ = 2.0%. Moreover, also GGMA at $C_w$ = 2.5% with the addition of $Ac_2Ca$ ($C_{Ac_2Ca}$ = 5.0 mM) was prepared. The reason for this choice was to obtain a GGMA sample with rheological characteristics similar to pure GG at $C_w$ = 2.0%, chosen as a reference, as described in more detail by ref. [30].

## 2.3. Microgel preparation

The preparation protocol of gellan gum microgel described in ref. [25] has been improved thanks to the employment of an Anton Paar MCR102 rheometer (Anton Paar Group AG, Graz, Austria). Unlike the previous method, where the GG solution was prepared in a beaker with the help of a magnetic stirrer and heated in a water bath using a manually controlled hot plate (aiming to maintain the raising temperature rate of 0.5 °C/min), the preparation process is now automated using a rotational rheometer. This ensures a more precise and consistent temperature and stirring control during the preparation, that is fundamental for the for sample preparation reproducibility ensuring that gelation occurs in the exact same way. Pure GG and GGMA microgels were prepared dissolving the powder in ultrapure water at room temperature under stirring as also described in ref. [22, 25]. Then the solution was poured in a cylindrical cuvette, placed on an Anton Paar rheometer and heated up to T = 80 °C, under the application of a constant shear rate of 500 s$^{-1}$ and left for 10 minutes in this condition. An opportune aliquot of concentrated NaCl solution (1 M) was added to obtain the chosen salt concentration $C_s$ and promoting the gelation process. Then, the solution was cooled from T = 80 °C to T = 25 °C at a cooling rate of 0.5 °C/min keeping the shear rate constant through the rheometer. GG and GGMA microgels were prepared at $C_w$ = 0.1%, $C_w$ = 0.2%, 0.3%, 0.5% at fixed NaCl concentration $C_{NaCl}$ = 100 mM.

## 2.4. Rheological Measurements

Rheological measurements were carried out with a rotational rheometer Anton Paar MCR102 with a cone plate geometry (plate diameter =

49.97 mm, cone angle = 2.006°, truncation = 212 μm). Temperature was controlled using a Peltier system equipped with an evaporation blocker and an isolation hood to prevent evaporation. Rheological measurements on GG and GGMA hydrogels and microgels at different weight and salt concentrations were performed both in oscillatory and in steady shear regimes. Amplitude sweep measurements of $G'(\gamma)$ and $G''(\gamma)$ vs shear strain $\gamma$ were used to determine the extent of the linear viscoelastic region (LVR) and the critical onset of non linearity. They were carried out in the strain range $\gamma$ = (0.01-100)% at fixed frequency. Frequency sweep tests of $G'(\omega)$ and $G''(\omega)$ vs frequency $f = \omega/2\pi$ were performed in the frequency range $f$ = (0.01-100) Hz at an applied strain $\gamma$ small enough to ensure that it is in linear viscoelastic regime where most of viscoelastic materials behave linearly with direct proportion-ality between the stress $\sigma$ and the deformation $\gamma$ [41]. These measurements were performed to assess the effects on the mechanical spectra. Temperature sweeps of $G'(T)$ and $G''(T)$ vs temperature were carried out to asses the sol-gel transition of both GG hydrogels and microgels in the temperature range (25-100) °C depending on the sample.

*2.5. Dynamic Light Scattering (DLS) Measurements*

The particle hydrodynamic radius $R_h$ has been measured through Dynamic Light Scattering (DLS) as a function of temperature. An optical setup based on a solid state laser (100 mW) with monochromatic wavelength $\lambda$ = 642 nm and polarized beam has been used to probe samples in dilute regime. Measurements have been performed at a scattering angle $\vartheta$ = 90° that corresponds to a scattering vector Q = $(4\pi n/\lambda) \sin(\vartheta/2)$ = 0.018 nm$^{-1}$ where n=1.33 is the solvent (water) refractive index. The hydrodynamic radii have been obtained through the Stokes-Einstein relation:

$$R_h = k_B T/6\pi\eta_s D_t \quad (1)$$

where $k_B$ is the Boltzmann constant, $\eta_s$ the viscosity of the solvent, namely water, at the measured temperature and $D_t$ the translational diffusion coefficient related to the relaxation time $\tau$ through the relation:

$$\tau = 1/(Q^2 D_t). \quad (2)$$

The relaxation time was obtained by fitting the autocorrelation function of scattered intensity through the Kohlrausch-William-Watts expression [42, 43]:

$$g_2(Q, t) = 1 + b[exp(-(t/\tau)^\beta)]^2 \quad (3)$$

with the stretching exponent β providing the deviation from the single exponential. The decay of the autocorrelation function provides insights into the diffusion behavior of microgels. A slower decay implies a longer relaxation time, showing slower dynamics due to increased interaction and frictional forces from larger hydrodynamic radii. Conversely, a faster decay indicates smaller size and quicker diffusion rates.

To ensure a constant temperature, the samples were kept at T=25 °C for 5 minutes before measurements.

## 3. Results

*3.1. Hydrogels rheological characterization*

*3.1.1. Pure GG hydrogels characterization*

We start by reporting the rheological behavior of pure GG hydrogels at different weight concentrations and salt content with the aim to understand the viscoelastic properties as the polysaccharide content increases. Although this problem has been investigated before, the weight concentration of GG has not reached large enough values as required by the use for paper cleaning in the cultural heritage field [44].

We start by reporting the rheological behavior of pure GG hydrogels at different weight concentrations and salt content with the aim to understand the viscoelastic properties as the polysaccharide content increases. Although this problem has been investigated before, the weight concentration of GG has not reached large enough values as required by the use for paper cleaning in the cultural heritage field [44]. Amplitude sweep experiments were carried out to estimate the dynamic linear viscoelastic range in oscillatory shear strain. Fig. 1(a) illustrates the storage, $G'(\gamma)$, and loss, $G''(\gamma)$, moduli of pure GG hydrogels at different concentrations, at f = 1 Hz and T = 25 °C, as a function of shear strain $\gamma$. At the lowest concentration, $C_w$ = 0.5%, $G''(\gamma)$ is greater than $G'(\gamma)$ over the entire strain range, typical behaviour of a liquid like system. With increasing concentration the trend is reverted and at approximately $C_w$ = 1.0%, $G'(\gamma)$ becomes greater than $G''(\gamma)$, followed at larger concentrations by an increase of the moduli of several orders of magnitude, indicating the onset of solid like behaviour. Interestingly, the onset of non-linear response is detected at the highest studied concentrations, i.e., $C_w$ = 3.0% and 4.0%. In particular, $G''(\gamma)$ displays a peak whose maxi- mum corresponds to the inversion point of the moduli and to a decrease of $G'(\gamma)$. This occurs at the so-called breaking point, characterized by the yield strain $\gamma_c$ (arrows in Fig. 1(a)), which is found to shift to smaller values as $C_w$ increases, suggesting the presence of a more rigid and brittle structure. To define $\gamma_c$, we use the intersection between $G'(\gamma)$ and $G''(\gamma)$, following literature. Recently, the increase of the loss modulus at the onset of non-linear dynamic response has been related to energy dissipation associated to microstructure reorganization before collapsing under shear [15, 45] and also qualitatively connected to the strength of the network in response to increasing deformation [15, 46]. Additional measurements are reported in the SI.

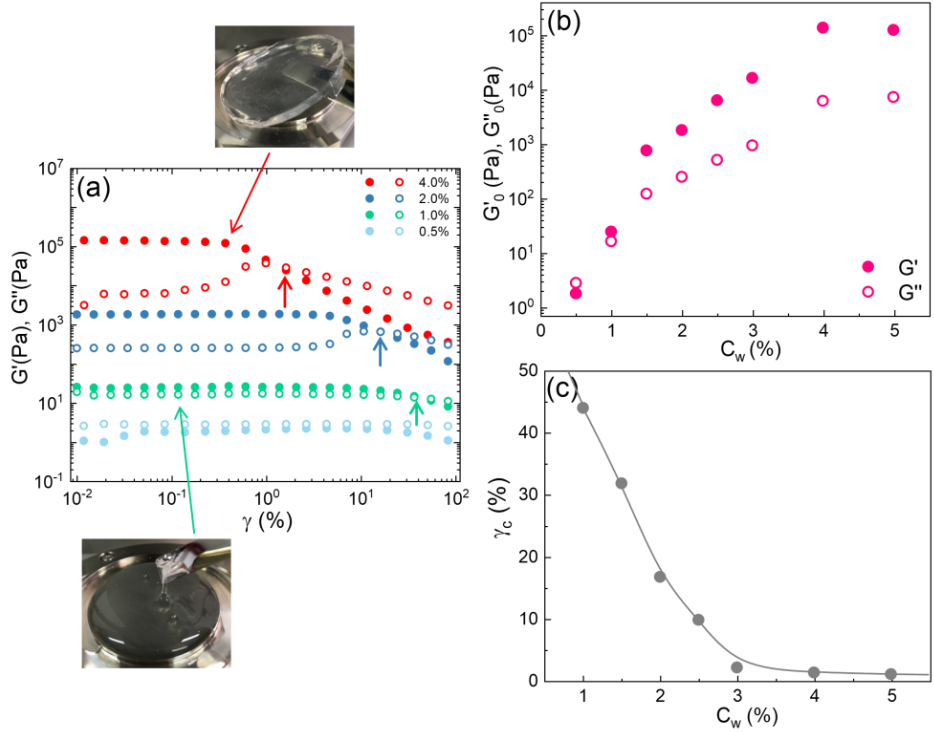

Figure 1: (a) Storage $G'(\gamma)$ (closed circles) and loss $G''(\gamma)$ (open circles) moduli as a function of shear strain $\gamma$, at f = 1 Hz and T = 25 °C for pure GG hydrogels at four different concentrations (0.5% light blue, 1.0% green, 2.0% blue and 4.0% red), with related photographs of GG in the fluid and in the solid states. Short arrows indicate the breaking point $\gamma_c$. (b) Plateau moduli $G'_0(\gamma)$ and $G''_0(\gamma)$ and (c) breaking point $\gamma_c$ as a function of GG concentration.

Plateau moduli $G'_0(\gamma)$ and $G''_0(\gamma)$, obtained at low strain $\gamma = 0.1\%$ in the linear viscoelastic region, are reported in Fig.1(b) as a function of concentra-tion, showing an increasing trend with $C_w$, in agreement with previous studies [47, 48] performed in a concentration range from 0.005% to 0.05%, albeit significantly lower than the present one. The breaking point $\gamma_c$ as a function of concentration, in Fig. 1(c), decreases with increasing concentration. The trend is well described by a decreasing exponential, reaching a plateau value $\gamma_c$ =(1.8-1.9)% at high concentrations. This indicates that from a threshold concentration value of about 3.0% the microscopic polysaccharide structure has similar brittle properties. The figure showing the moduli as a function of frequency is provided in the SI and is consistent with the behavior of the amplitude sweep, confirming our observations. At $C_w$ = 1.0%, the system is in a fluid state as shown in Fig.1 (a); however $G'(\gamma) > G''(\gamma)$ over the entire frequency range indicating a dominant elastic behaviour. At concentrations $C_w$ = 2.0% and 3.0%, $G'(\gamma)$ and $G''(\gamma)$ are almost independent on frequency and $G'(\gamma)$ is higher than $G''(\gamma)$, typical hallmark of a solid like behaviour. The observed trend with concentration supports the increase of gel domains with increasing concentration, in agreement with what reported in ref. [15], in which, however, only very low concentrations from 0.025% to 0.25% were investigated.

*3.1.2. Sol-gel transition of GG hydrogels*

As mentioned above, the gelation process of GG is a complex phenomenon influenced by many factors, including the physicochemical properties of the solvent, temperature, and ionic concentration. In physical gels such as the present ones, an accurate detection of the sol-gel transition can be experimentally challenging. Gellan gum gelation is commonly described as a two-step thermoreversible process with an initial helix ordering, followed by the association between stiff double helices through intermolecular interactions [49, 31], recently confirmed by molecular dynamics simulations [34]. Typically, gela- tion in GG systems occurs by decreasing temperature or by increasing ionic strength [50]. Here, we first focus on the temperature behavior by perform- ing dynamic moduli measurements versus temperature during gelation, as reported in refs. [51, 52, 53, 50]. The gel point temperature, $T_c$, is defined as the temperature where the moduli cross over. According to it, by monitoringthe temperature evolution of the moduli during cooling at fixed frequency(f = 1 Hz, $\omega$ =6.28 rad/s), we can obtain an estimate of the gel point asthe crossover point of $G'(T)$ and $G''(T)$, corresponding to tan$\delta$ = 1, later followed by a trend inversion whereby $G''(T) > G'(T)$. In Fig. 2(a) the tem- perature dependence of dynamic moduli and tan$\delta$ during the cooling process (2.0 °C/min) is reported for GG samples at concentrations: $C_w$ = 1.0%, 2.0% and 4.0% highlighting the onset of gelation. At the lowest concentra- tion, 1.0%, gelation occurs for $T_c$ 80 °C while at 2.0% this occurs upon cooling already approximately 93 °C. Further increasing GG concentration to 4.0%, no intersection between dynamic moduli is detected, with $G'(T)$ always being larger than $G''(T)$ throughout all the cooling range. This indi- cates that gelation in this case is instantaneous as soon as the heating processis stopped and the sample start to cool. An increase of the plateau of the moduli can be observed when temperature further decreases below the gel point (T < $T_c$). This trend may be due to the fact that the aggregation of double helices into the ordered conformation of GG is not energetically pre-ferred at temperatures just below $T_c$, while lower temperatures enhance their formation. Fig.2(b) reports the moduli, together with their respective tan$\delta$ for the sample at C$_w$=2.0% to highlight that they overlap in correspondence of the gel point where tan$\delta$ = 1.

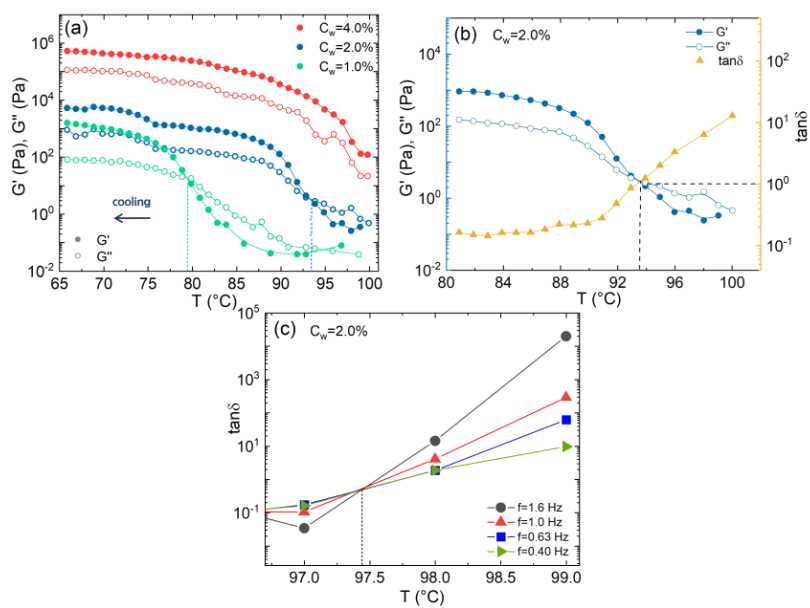

Figure 2: (a) Temperature evolution of $G'(T)$ (closed cyrcles) and $G''(T)$ (open cyrcles) during cooling for pure GG at a rate of 2.0 °C/min, at f = 1 Hz and at $C_w$=1.0% (green), 2.0% (blue) and 4.0% (red). (b) Temperature dependence of the moduli and tan$\delta$ (orange triangles) for pure GG at $C_w$=2.0%. The crossover point defines $T_c$ = 93.6 °C. (c) Temperature dependence of tan$\delta$ for pure GG at different angular frequencies: f = 1.6 Hz (grey circles), f = 1.0 Hz (red triangles), f = 0.63 Hz (blue squares) and f = 0.40 Hz (green triangles). The crossover point defines $T_{gel}$ = 97.4 °C according to the critical phase angle procedure [54].

Nonetheless, it has been argued that the temperature corresponding to the crossover, T$_c$, may vary depending on the frequency [50] at which moduli are measured and therefore it cannot be really considered an effective sol-gel transition temperature, T$_{gel}$, although very close to T$_{gel}$. Following the work by Winter and Chambon [54], an interpolation method can be implemented, known as the critical phase angle procedure where tan$\delta$ is frequency independent. The convergence point of the tangent at various angular frequencies determines the gelling temperature, T$_{gel}$. In Fig. 2(c) the temperature dependence of tan$\delta$ of pure GG is shown for different angular frequencies at C$_w$= 2.0%. The convergence point of all frequencies is found to be at T$_{gel}$ = 97.4 °C, corresponding to the point where the critical phase angle $\delta$ becomes frequency independent. This value is slightly greater, but quite close to T$_c$ = 93.6 °C, so that the latter can be considered as a good approximation to the gel point. Overall, these measurements indicate that the best suitable GG hydrogel without salts for cultural heritage purpose is that with C$_w$=4.0%: it could indeed withstand a pressure up to 1000 Pa without breaks (see Fig. 6).

*3.1.3. Characterization of GG hydrogels in presence of salts*

The gelation mechanism of GG has been widely discussed in literature [19, 20, 55, 34]. It is assumed that, in the second step of gelation, the presence of salts promotes the gel formation affecting the double helix structure due to electrostatic interactions with cations. In particular, the mechanical properties of GG in presence of different salts have suggested that the aggregation of the double helices follows distinct mechanisms in the presence of monovalent or divalent cations [34]. For divalent ions at a given ionic strength, GG hydrogels are firmer and harder with respect to those prepared at the same ionic strength but with monovalent ions [48, 18]. Here, we complement this evidence for the gels used for paper cleaning applications. In particular, we carry out rheological measurements adding to GG different amounts of NaCl and Ac$_2$Ca salts to form hydrogels. These are compared to the correspond-ing results for pure GG, keeping its concentration constant at C$_w$ = 2.0%, that is the one that was found to be best performing in the paper cleaningtests [29].

In Fig. 3, we report G$'$($\gamma$) and G$''$($\gamma$) vs $\gamma$ for different GG hydrogels prepared in the presence of varying divalent, salt $Ac_2Ca$ (Fig. 3(a)(c)), and monovalent one, NaCl, amount (Fig. 3(b)(d)). The addition of salts leads, in both cases, to an increase of G′(γ) of one order of magnitude with respect to pure GG hydrogel as shown in Fig. 3 (a) (b); however the behavior of hydrogels are quite different for the two cases. Indeed, in the presence of divalent cations, with increasing $Ac_2Ca$ concentration, a continuous increase of G′(γ) is observed, while for NaCl a sudden change with respect to pure GG is obtained at the lowest investigated salt concentration, without further increase even by at the highest NaCl investigated concentration. This is in qualitative agreement with molecular dynamics simulations, which pointed out that divalent salt is able to efficiently bridge GG chains into aggregated double helices, while monovalent ions cannot act as effective linking agents, but can only increase the screening between the GG chains. Therefore, their role is not prominent and cannot be used as a control way to obtain hydrogels of the desired stiffness. Similar results are observed for the loss moduli $G''(\gamma)$ in Fig. 3 (c) (d), where a peak that is larger and shifted to lower strain, is found with increasing Ac$_2$Ca content, while the corresponding behavior for NaCl is independent of the actual salt concentration. An additional compar-ison is reported in Fig.S2, where $G'(\gamma)$ and $G''(\gamma)$ at fixed salt concentration (2.5 mM) of NaCl and Ac$_2$Ca are compared with those of pure GG hydro-gel. Altogether, these results confirm that the viscoelastic behaviour of GG hydrogels is influenced much more strongly by divalent cations than by monovalent cations, as reported in ref. [18] , where the authors compared the effect of KCl, NaCl, CaCl$_2$ and MgCl$_2$ concluding that divalent cations promote the formation of thermally stable junction zones. To better visualize this effect, the plateau moduli at low strain values are shown as a function of salt concentrations $C_{salt}$ for GG with Ac$_2$Ca (red circles) and NaCl (orange squares) in Fig. 4(a). Furthermore, in Fig. 4(b) the critical strain $\gamma_c$ is re- ported, showing a smooth decrease with increasing salt concentration for GG in the presence of Ac$_2$Ca, which favors closer and tighter bindings between the chains at variance with NaCl. Finally for this section, we report in Fig. 5 the comparison of $G'(\gamma)$ and $G''(\gamma)$ at $C_w = 2.0\%$ for pure GG hydrogel andfor GG hydrogels with different concentrations of Ac$_2$Ca and NaCl, but in such a way to keep constant the ions number (normality) in the different samples. Figs. 5(a)(d) show that a very similar sample response is obtained at low salt content, despite using different monovalent or divalent salts. Interestingly, in the presence of a relatively high salt content, we find a quite pronounced variation of the moduli, as shown in Fig. 5(c)(f). It should be noted that in literature the GG hydrogels with salts used for cultural heritage purpose contains $Ac_2Ca$ = 2.50 mM, and therefore fall in this last category. The effect of divalent cations is further revealed by looking at the stress reported in Fig. 6, where we compare shear stress value $\sigma$ as a function of shear strain $\gamma$ of the hydrogels without added salt (a), with calcium ions (b) and with sodium ions (c), separately. Strikingly, we find that in the presence of divalent ions, evidence of two-step yielding behaviour is observed, right at the conditions where we use these GG hydrogels for paper cleaning. Two-step yielding is associated with the occurrence of the breaking of the gel in two distinct processes [35]. Here, we can hypothesize that at first the breaking of the cation-mediated aggregation of double helices occurs, and then the breaking of the double helixes themselves. The two steps are separated by a difference in stress of roughly one order of magnitude (from about $\sigma$=110 Pa in the first step to $\sigma$=1100 Pa in the second step). Our cleaning tests, reported in literature [29, 28], definitely exclude that the gel leaves residues on paper, when pressures of the order a few hundreds Pa are exerted, while residues can remain on paper if pressure of thousands of Pa are put on the gel. Therefore, we deduce that gel is able to perform its cleaning function up to the second yielding point, with the presence of divalent cations giving it the necessary rigidity to be removed afterwards without any leaving residues. Instead, we do not observe double yielding in the presence of monovalent salt at the same ions number, which confirm the fact that these cannot be used for cultural heritage purposes, since they are too soft, they can break after the weigth is applied to it, leaving residues on paper sheets after process. Interestingly, double yielding also appears for pure GG hydrogel at higher $C_w$, about 4.0% signaling that the presence of two-step yielding is intrinsic in the complex gelation process of GG, again depending on the particular conditions employed.

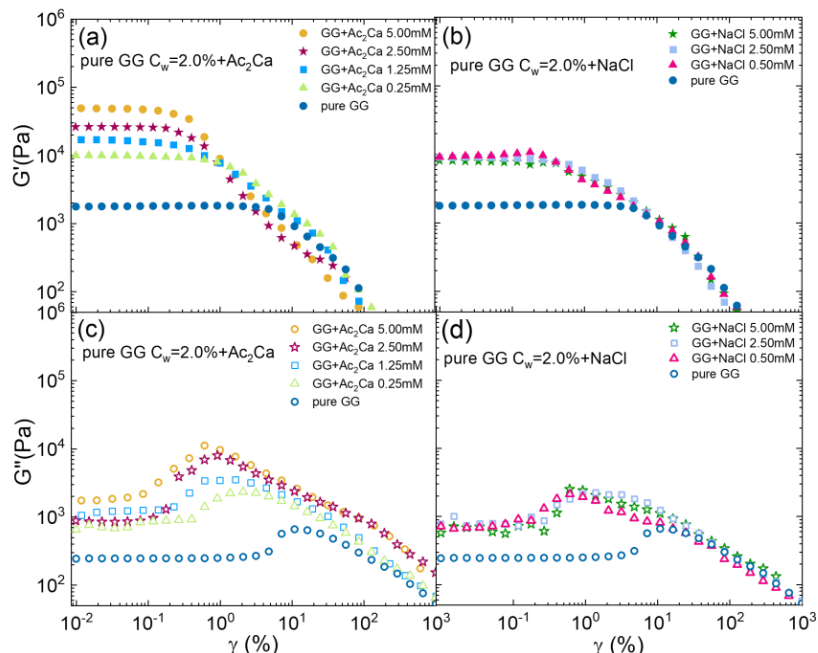

Figure 3: (a) (b) Storage G′(γ) (closed circles) and (c) (d) loss G″(γ) (open circles) moduli as a function of shear strain γ at T = 25 °C at $C_w$ = 2.0% with different $Ac_2Ca$ contents (0.25 mM (light green triangles), 1.25 mM (light blue squares), 2.50 mM (purple stars) and 5.00 mM (orange circles)) or NaCl contents (0.500 mM (pink triangles), 2.50 mM (light violet squares), and 5.00 mM (green stars)), compared with pure GG (blue circles).

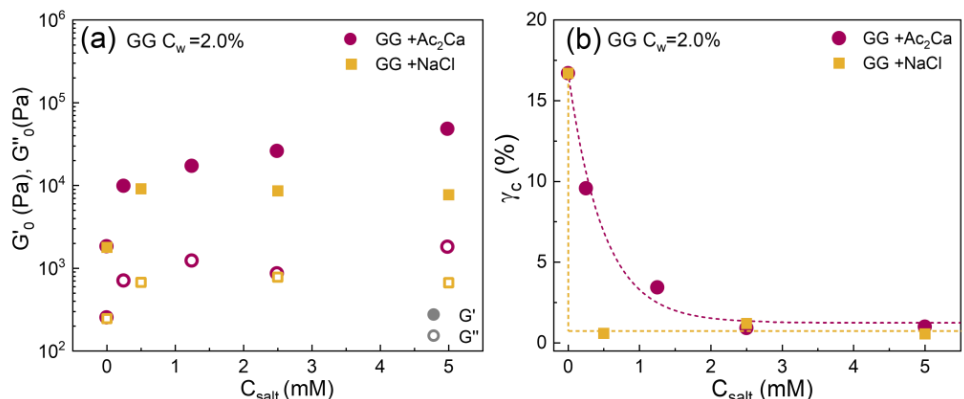

Figure 4: (a) Plateau moduli $G'_0(\gamma)$ and $G''_0(\gamma)$ at $\gamma = 0.1\%$ and (b) breaking point $\gamma_c$ as a function of $C_{salt}$ for GG hydrogels at $C_w = 2.0\%$ with Ac$_2$Ca salt (red circles) and NaCl (orange squares). Lines are guides to the eye.

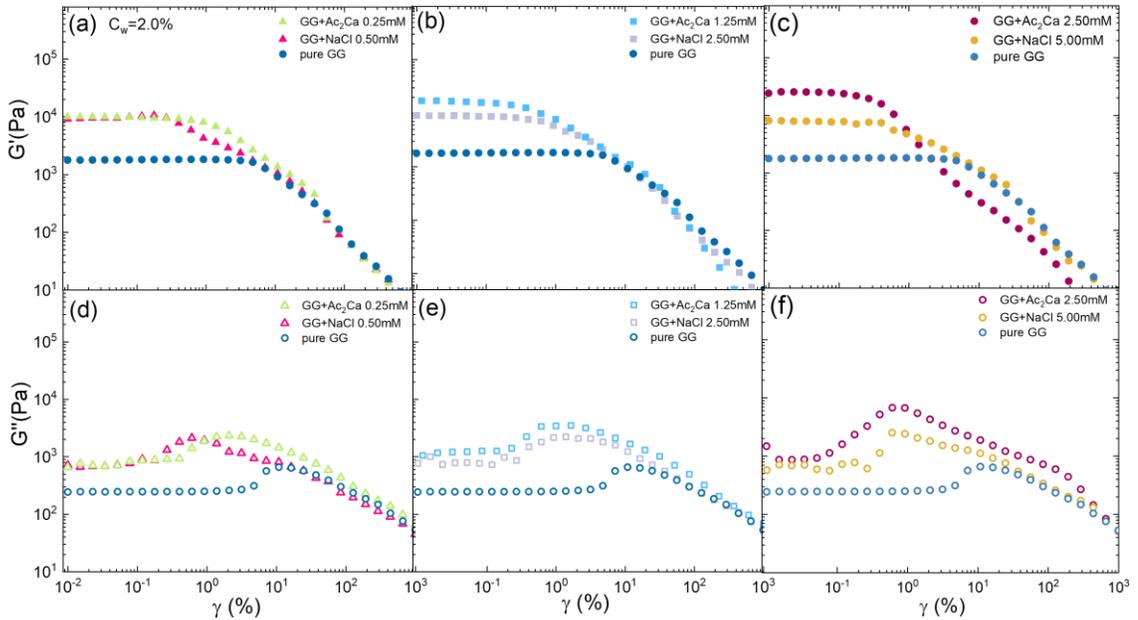

Figure 5: (a) (b) (c) Storage $G'(\gamma)$ and (d) (e) (f) loss $G''(\gamma)$ moduli vs shear strain $\gamma$ at T = 25 °C and $C_w$ = 2.0% for pure GG hydrogel and for GG hydrogel with $Ac_2Ca$ and NaCl at salt concentrations such as to keep constant the ions number in the sample (0.25 mM, 0.50 mM, 1.25 mM, 2.50 mM and 5.00 mM).

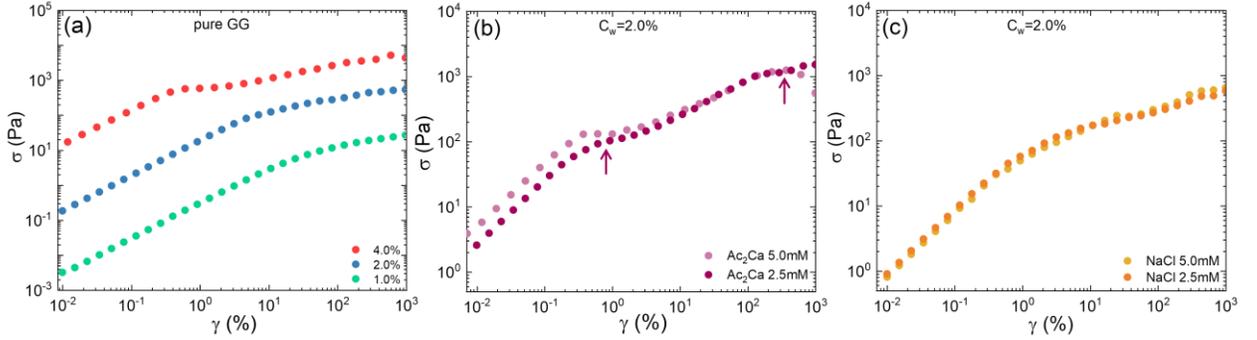

Figure 6: Shear stress $\sigma$ as a function of shear strain $\gamma$, at f = 1 Hz, T = 25 °C for (a) pure GG hydrogel at different concentrations and for GG hydrogels at $C_w$ = 2.0% (b) with $Ac_2Ca$ and (c) NaCl added salts. The arrows in (b) indicate the double yielding for the sample which best performs in paper cleaning.

### 3.1.4. Sol-gel transition of GG hydrogels with cations

It is interesting analyze the sol-gel transition of GG in the presence of salts. As discussed above, the addition of cations (monovalent or divelent ones) largely affects the viscoelastic behaviour of GG, as evident from the increase of the moduli by some orders of magnitude. This suggests that, in this condition, gelation should occur at lower GG concentrations in the presence of cations. Based on these premises, we investigate the gelation process of 0.5% GG samples with the addition of $Ac_2Ca$ and NaCl salts at the same ions number cases. In Fig. 7, G′($T$ ) and G″($T$ ) versus temperature during the gelation process at cooling rate 2.0 ° C/min and frequency f = 1 Hz are reported at $C_w$ = 0.5% for pure GG, GG with $Ac_2Ca$ 2.5 mM and GG with NaCl 5.0 mM. First, we observe that at this low concentration, as discussed in the related section, pure GG remains liquid in the whole investigated temperature range, as evident from the low values of the moduli, whereas the addition of salts promotes gelation already at this low $C_w$. Interestingly, we observe that in the case of added $Ac_2Ca$, the inversion of the moduli occurs at slightly higher temperature, T = 46.2 °C, as compared to GG with NaCl, T = 40.5 °C, signaling that upon cooling the divalent ions are more effective in inducing gelation.

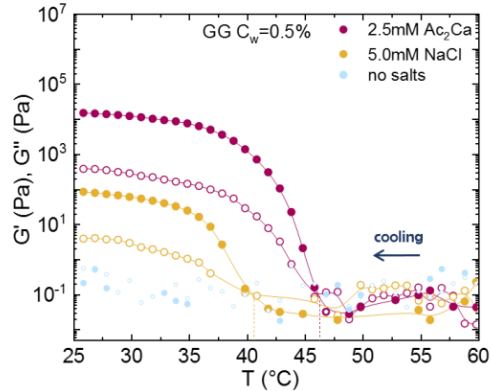

Figure 7: Temperature evolution of G′($T$ ) (closed symbols) and G″($T$ ) (open symbols) during cooling at rate 2.0 °C/min at frequency 1 Hz of GG at $C_w$=0.5% without salts compared with GG with 2.5 mM of $Ac_2Ca$ and 5.0 mM of NaCl. The crossover point defines T(GG with NaCl) = 40.5 °C and T(GG with $Ac_2Ca$ =46.2 °C.

Importantly, low temperature plateaus of G′($\gamma$) and G″($\gamma$) increase by one and two orders of magnitude with respect to pure GG in the case of NaCl and $Ac_2Ca$, respectively. These findings further confirm the prominent role of cations in general and of divalent ones in the viscoelastic behaviour of GG hydrogels [18, 34] .

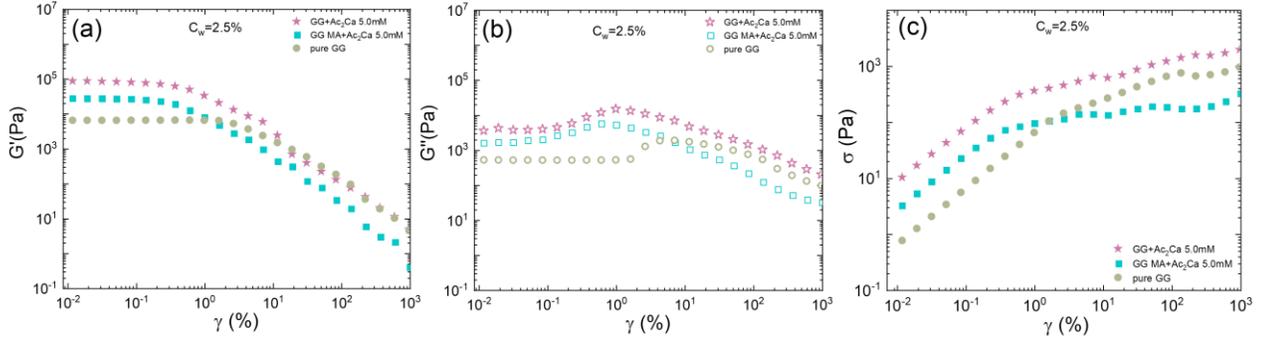

Figure 8: (a) G'(γ), (b) G''(γ) and (c) σ vs shear strain γ at T = 25 °C and $C_w$ = 2.5% for pure GG hydrogel, GG hydrogel with 5.0 mM of $Ac_2Ca$ and methacrylate GG hydrogel with the same content of salt.

*3.1.5. Methacrylate gellan gum (GGMA) hydrogels*

Finally, we focus on the chemical modification to gellan gum by methacrylation. This is because, the introduction of methacrylic groups allows to obtain hydrophobic gel, yielding a gel that is capable of removing both hydrophilic (i.e. cellulose degradation products) and hydrophobic substances with a single water-based treatment [30], a novel feature in paper cleaning procedures. In this case, based on results previously reported concerning the best hydrogels for paper cleaning, we focus here on methacrylated GG with the addition of $Ac_2Ca$ only, having established in the previous section that this salts yields superior gels with respect to NaCl.

To characterize the rheological properties of these chemically-modified hydrogels, we report in Fig. 8 a comparison among pure GG, GG with 5.0 mM of $Ac_2Ca$ and methacrylated gellan gum (GGMA) with the same content of salt and polymer concentration of $C_w$ = 2.5%. Also in the case of GGMA, the addition of $Ac_2Ca$ causes an increase of G'(γ), suggesting a more tight structure between the chains and a more rigid gel. The slight difference in used $C_w$ is due to the fact that after various attempts described in ref. [26], we found that a slightly larger amount of polymer is needed in the case of GGMA, i.e., $C_w$ = 2.5% (versus $C_w$ = 2.0% for pure GG) to obtain the same optimal hydrogel performance for paper cleaning. This is confirmed by the lower elastic moduli observed for GGMA with respect to the corresponding pure GG in Fig. 8. The trend of the plateau moduli $G'_0(γ)$ and $G''_0(γ)$ and of the critical strain $γ_c$ with increasing GG concentration, with and without $Ac_2Ca$ and GGMA with the same salt is shown in Fig. 9. Notably, we find that the two hydrogel formulations which are found to best perform in paper cleaning, namely GG at $C_w$ = 2.0% with $Ac_2Ca$ 5 mM and GGMA at $C_w$ = 2.5% with $Ac_2Ca$ 5 mM, have really comparable elastic properties, as highlighted by the horizontal line in Fig. 9(a).

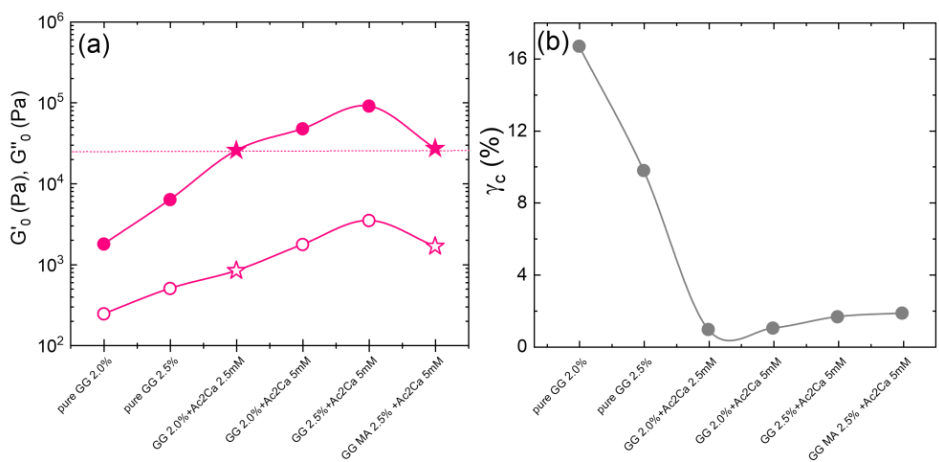

Figure 9: Comparison of (a) the plateau moduli $G'_0(\gamma)$ (closed circles) and $G''_0(\gamma)$ (open circles) and (b) $\gamma_c$ for the different GG and GGMA hydrogels tested, varying polymer and salt concentrations, at T = 25 °C. The horizontal line highlights the comparable value of $G'_0$ obtained for the two hydrogels which best perform in paper cleaning.

### 3.2. Microgels characterization
#### 3.2.1. GG microgels

Combining the promising features of microgels and the proven efficiency of GG hydrogels in removing impurities and degradation products from paper, GG microgels were developed with the aim of obtaining an innovative tool capable of improving the efficiency of currently available cleaning techniques for paper artworks [25].

As reported in ref. [25], GG microgels have been prepared and successfully employed, for the first time, in paper cleaning. Thanks to their reduced size, microgels have proven to be effective, providing a greater penetration into the porous structure of the paper adapting to the irregular surface of artefacts, with respect to hydrogels. Thanks to their properties, microgels can cleanvery quickly, in the order of a few minutes, thus strongly minimizing timecosts with respect to hydrogels (whose application can be up to two hours

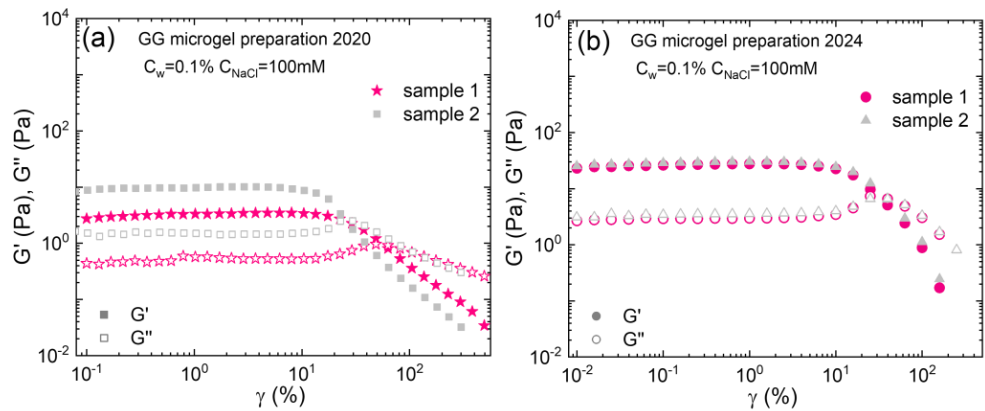

Figure 10: Comparison between amplitude sweep measurements (storage modulus G', closed symbols, and loss G'', open symbols vs shear strain $\gamma$), at f = 1 Hz, T = 25°C in the case of microgels at $C_w$=0.1% and $C_{NaCl}$ = 100 mM prepared by (a) the old method published in 2020 [25] and (b) the new rheometer-based method proposed in this work. In both panels two different samples (sample 1, pink colored and 2, grey colored) are compared for each method.

long).

We start this characterization by comparing the properties of GG microgels obtained with our initial preparation method, put forward in Ref. [25] and detailed in the Materials and Methods section, with those prepared in this work, thanks to the use of a rheometer, which allows for greater control over the produced samples. Fig. 10 shows the comparison of rheological measurements on the microgels obtained using the two methods. In Fig. 10(a) amplitude sweep results are conducted on two different sample preparations, both obtained with the initial preparation method. It is evident that, while the behavior of the two samples is qualitatively similar, there is some differences in the measured values of the moduli. Conversely, the measurements carried out on two different samples prepared using the new rheometer-based method, in Fig. 10(b), are completely overlapped, indicating a significant improvement in consistency and reproducibility of the samples. We have verified that, in both cases, samples have the same cleaning efficacy, because despite small differences, microgels are able to efficiently and rapidly penetrate into the pores of the fibrous network of paper sheets, due to their small size, as will be discussed better below. In the following, all microgel results refer to samples prepared with the new method, as described in the section Materials and Methods, at the following GG concentrations: $C_w = 0.1$% (sample GG1), 0.2% (sample GG2), 0.3% (sample GG3), 0.5% (sample GG5). We now an- alyze in detail the gelation process during microgel formation. In Fig. 11(a) the normalized viscosity of GG1 is reported as a function of temperature dur- ing the heating and cooling process under shear. During the initial heating, a quite linear decrease, most likely due to the chains forming double helix structures [55, 34], is found. However, when cooling starts, a sudden increase of viscosity is observed at a temperature close to 40°, indicating the start of gelation where the double-helixes form networked structures [19, 55].

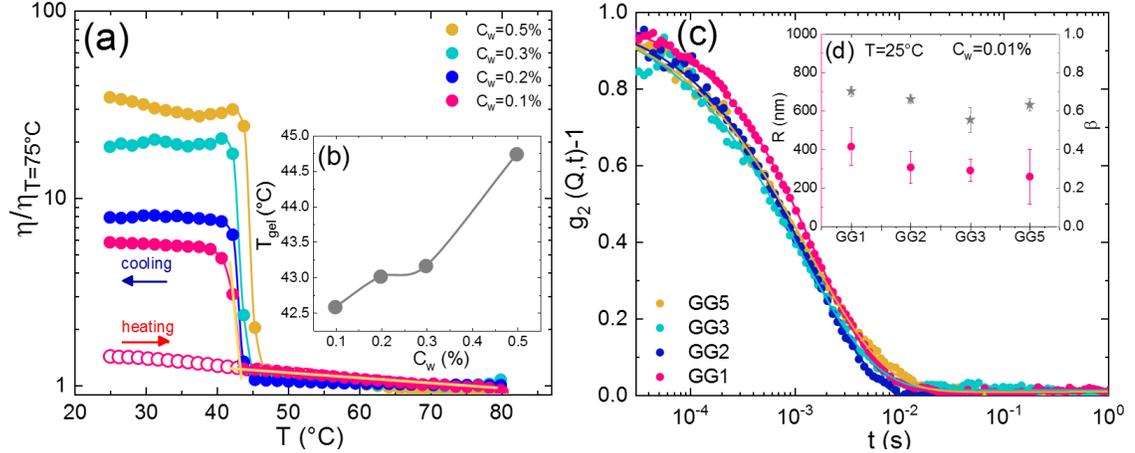

Figure 11: (a) Normalized viscosity vs temperature during the two steps of microgel preparation: during cooling (indicated by the blue arrow) for GG microgel at $C_w$=0.1%,0.2%,0.3% and 0.5% and during heating (indicated by the red arrow) for GG at $C_w$=0.1%, under a constant shear rate of $500\,s^{-1}$.
(b) Gelling temperature as a function of concentration, determined from the intersection point between the tangent at the base and at the midpoint height of the inflection of the curve (see orange curves drawn as an example for the $C_w$=0.1% sample). (c) Normalized intensity autocorrelation function $g_2(Q,t)-1$ of GG1, GG2, GG3 and GG5 samples (corresponding to microgels prepared by the new method at GG at $C_w$=0.1%,0.2%,0.3% and 0.5% ) measured in dilute condition, at $C_w = 0.01$%, and $\theta = 90°$, corresponding to $Q = 0.018\,nm^{-1}$. (d) Hydrodynamic radius R (red circles) and stretching parameter $\beta$ (grey stars) obtained from fits of $g_2(Q,t)$ through the eq. 3, (see Materials and Methods section).

In Fig. 11(a) the normalized viscosities vs temperature of the different GG

microgels samples during the cooling process are also shown. The viscosity plateau value at low temperature increases with increasing concentration and a shift of the gelling temperature to higher values is observed, suggesting the development of more rigid gel structures. Gelling temperatures as a function of GG concentration, reported in Fig. 11(b), show that, at fixed salt content, samples with higher polysaccharide content, reach the gel state earlier during the cooling process. The normalized DLS autocorrelation functions measured in dilute condition ($C_w$ = 0.01%) are reported in Fig. 11(c). They are directly correlated with the size of the particles through the Stokes-Einstein relation eq.1. In particular, the particle radii are shown in Fig. 11(d), together with the stretching parameter $β$. We find that the radii are always in the order of a few hundreds nanometer, with a slight decrease in size as GG concentration increases. In concurrence, there is a slight decrease of $β$, that is related to the polydispersity of the samples, that appear to be quite pronounced in comparison to standard microgel synthesized by precipitation polymerization or other techniques [56] . It is important to compare these results with those obtained with the old preparation method in ref. [25] , where the microgels where found to have a larger size (close to 1 $μ$m radius) and a lower stretching index, close to 0.4. Hence, both methods yield microgels capable to clean well the paper artefacts. This is because the pores of paper are also quite polydisperse and hence, the fine details are not crucial for their efficacy [57].· The key feature is truly the presence of a moderate polydispersity of the samples which allows them to better adapt to the rough paper surface.

To probe the viscoelastic properties of GGx (where x is 1,2,3,or 5) microgels, storage and loss moduli were measured at 25·C) as a function of shear strain and fixed frequency 1Hz and ii) as a function of frequency at fixed shear strain $γ$=0.1% in the linear viscoelastic region for all samples, as reported in Fig. 12(a) and (b), respectively. In Fig. 12(a), $G'(γ)$ is higher than $G''(γ)$ in the linear viscoelastic range, for all the samples, indicating that the elastic behavior prevails over the viscous one. As the concentra- tion increases, from GG1 to GG5, both $G'(γ)$ and $G''(γ)$ increase, suggesting a denser and more interconnected microgel structure. The breaking point, which marks the beginning of the non-linear response, shifts from 30%, for the sample with low concentration (GG1) to 3.0% for the more concentrated GG microgel (GG5) indicating that the microgel structure becomes more fragile as the concentration increases. In Fig. 12(b), $G'(f)$ remains higher than $G''(f)$ across the entire frequency spectrum, confirming the predomi- nantly elastic nature of the microgel network which exhibits weak frequency dependence over the tested range, a typical behavior for gel-like materials. The relative stability of the moduli across frequencies further indicates that the microgels maintain their structural integrity and viscoelastic properties, reinforcing their potential for applications requiring a consistent mechanical performance.

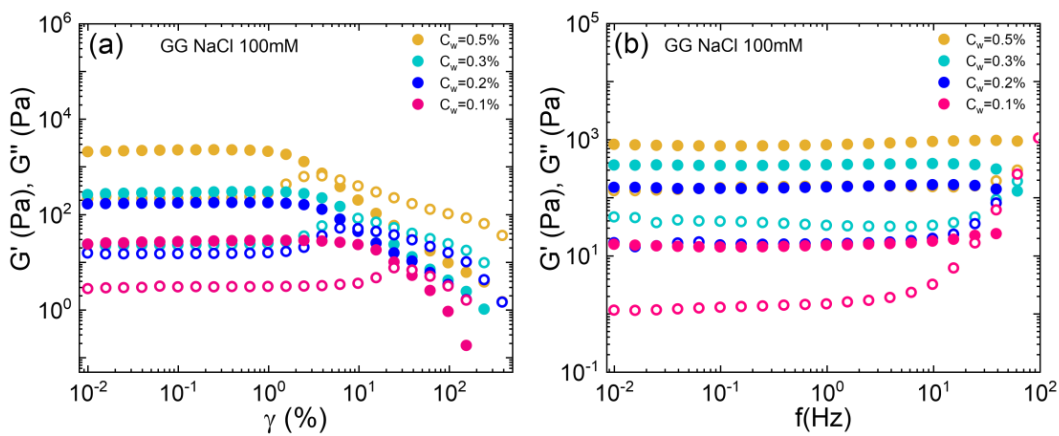

Figure 12: (a) Storage G′ (closed circles) and loss G″ (open circles) moduli as a function of $\gamma$ at f = 1 Hz and (b) as a function of frequency f at $\gamma$ = 0.1% for GG microgels at different concentrations, T = 25 °C and $C_{NaCl}$ = 100 mM.

*3.2.2. GGMA microgels*

The last preparation protocol was adopted for preparing GGMA microgels at different polymer concentrations, keeping fixed the NaCl content at 100 mM, in order to find the GGMA concentration for obtaining microgels with rheological properties similar to the successfully tested GG one ($C_w$ = 0.1%, $C_{NaCl}$ = 100 mM) [25] . This strategy ensures similar gel properties and interactions with paper surface. We thus tested the following GGMA concentrations: $C_w$ = 0.1% (sample GGMA1), 0.2% ( sample GGMA2), 0.3% (sample GGMA3), 0.5% (sample GGMA5).

In Fig. 13(a) we report the normalized viscosity vs temperature of the different samples during the cooling process. We observe a pronounced increase of viscosity with increasing GGMA concentration and a shift of the gela- tion to higher temperature, suggesting again the development of more rigid gel structures at higher polymer concentrations. Gelling temperatures as a function of GGMA concentration are reported in Fig. 13(b), showing that, similarly to pure GG microgels, samples with larger polysaccharide concen-tration and the same salt content reach, on cooling, the gel state first. The particle radii of the various GGMA samples, obtained by the DLS normal- ized autocorrelation functions, reported in Fig. 13(c) and in Fig.14(d), are comparable to those observed for GG samples, also showing similar stretch-ing index. Overall, these findings strongly support the idea that the adopted metachrylation procedure does not fundamentally alter the microgel struc-ture, just providing an additional functionality which makes them suitable to interact more efficiently with hydrophobic compounds.

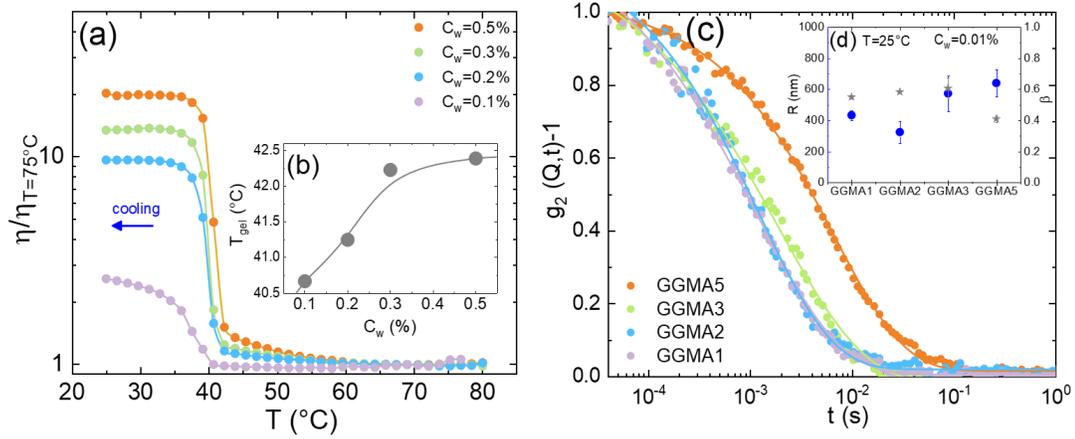

Figure 13: (a) Normalized viscosity vs temperature during cooling (blue arrow) for GGMA microgels at $C_w$=0.1%, 0.2%, 0.3% and 0.5% (called GGMA1, GGMA2, GGMA3 and GGMA5 repsctively), under a constant shear rate of 500 s$^{-1}$. (b) Gelling temperature as a function of concentration. (c) Normal- ized intensity autocorrelation functions of GGMA1, GGMA2, GGMA3 and GGMA5 samples measured in dilute condition, at $C_w$ = 0.01%, and $\theta$ = 90°, corresponding to Q = 0.018 nm$^{-1}$. (d) Hydrodynamic radius and $\beta$ obtained from fits to $g_2(Q,t)$ through the eq. 3, (see matherials and methods).

Next, we turn to discuss the storage and loss moduli vs shear strain, that are reported in Fig. 14 at f = 1 Hz and T = 25°C for GGMA at different concentrations. Again, we find that $G'(\gamma)$ is greater than $G''(\gamma)$ for all sam- ples and the plateau values increase with concentration, as in the case of pure GG microgels. There is a shift of the breaking point $\gamma_c$ from about 40%, for the less concentrated sample (GGMA1), to 30%, for the most con- centrated sample (GGMA5). The displacement to lower $\gamma$ of the beginning of the non-linear viscoelasticity zone with increasing polysaccharide concen-tration happens within a smaller range than for pure GG, suggesting that methacrylation makes it less susceptible to concentration changes. The sta- bility of the sample over time was tested by comparing the amplitude sweep measurements of freshly prepared samples with those several hours after the preparation, as reported in Fig.S3(a) and (b) for two different concentrations. Comparing the values of the moduli with pure GG, we find that the sample with $C_w$ = 0.2% GGMA is the most similar to $C_w$ = 0.1% GG. Indeed, this sample was tested for paper cleaning in ref. [26] and its efficacy was found to be the same as that for pure GG microgels, but with the additional ability to remove hydrophobic residues. Finally, we compare the flow curves of the microgels in Fig. 15 , for microgels (a) and for GGMA microgels (b), both as a function of polysaccharide concentration. Interestingly, despite the lower overall moduli with respect to the hydrogels, we also observe for microgels the emergence of double yielding, signaling that, independently of the internal arrangement in a macroscale network or in microspheres, the gelation mech- anism is the same. Here, it is peculiar that the microgels are obtained in the presence of monovalent ions, for which hydrogels did not show the two-step yielding, pointing to a difference induced by the shear-induced aggregation which seems to stabilize the micro-aggregates to a greater extent, making them more resistant to fracture. It is important to note that despite the low values of the stress observed here, the microgel-based cleaning protocol does not need the application of a weigth and involve a removal step using water. This effectively limits the production of residues and makes them effectivefor the desired purposes.

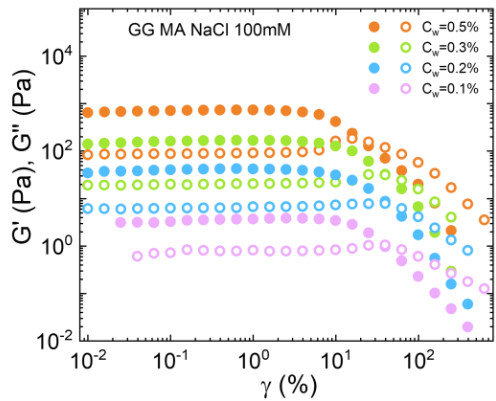

Figure 14: Storage G′ (closed circles) and loss G″ (open circles) moduli as a function of shear strain $\gamma$, at f = 1 Hz, T = 25 °C and $C_{NaCl}$ = 100 mM for GGMA microgels at different concentrations.

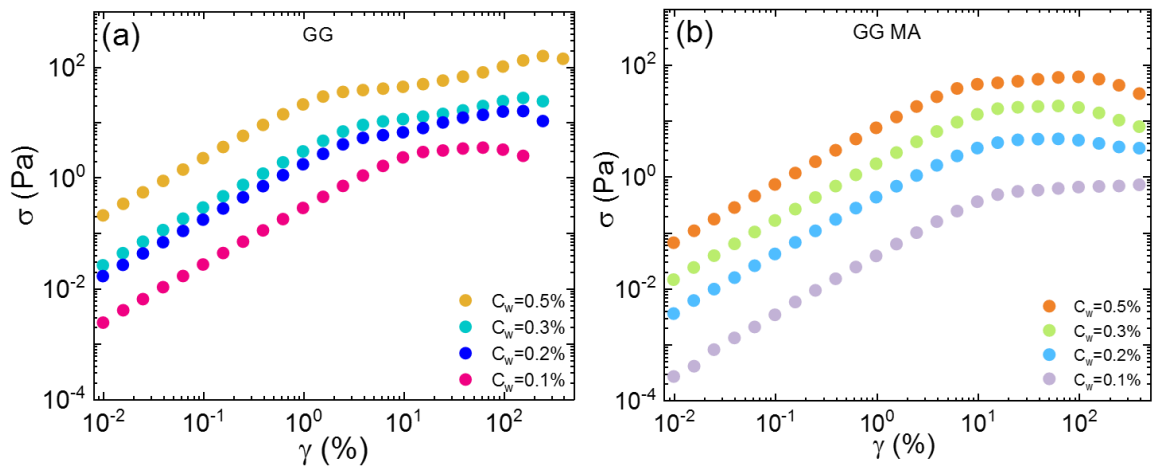

Figure 15: Shear stress $\sigma$ as a function of shear strain $\gamma$, at f = 1 Hz, T = 25 °C and $C_{NaCl}$ = 100 mM for (a) pure GG and for (b) GGMA microgels at different concentrations with fixed $C_{NaCl}$ = 100 mM.

## 4. Conclusions

In this work the rheological properties of GG hydrogel and microgels recently employed in paper cleaning restoration interventions have been thoroughly explored as a function of polysaccharide concentration, content and type of added salts (monovalent or divalent ones). With increasing polysaccharide concentration a broad increase of dynamical moduli with $G'(\gamma) > G''(\gamma)$ and a decrease of the critical deformation $\gamma_c$ are observed, suggesting the formation of more rigid and brittle structures (Fig. 1). The gelation mechanism is also deeply affected by the presence of cations (salt) which act on the double helix structure through electrostatic interactions promoting gel formation. In particular, the addition of divalent cations, at a given ionic strength, gives rise to GG hydrogels that are firmer at increasing polysaccharide content, $G'_{C_{w_1}} \gg G'_{C_{w_2}}$ with $C_{w_1} > C_{w_2}$, (Fig. 3 and Fig. 4) and also harder with respect to monovalent ones, $G'_{Ac_2Ca} \gg G'_{NaCl}$. As also microscopically shown in recent molecular dynamics simulations [34], this is due to the fact that divalent cations act as bridging point between double helices, simulations [34], thus promoting the formation of thermally stable junctions and a more rigid gel structure (Fig. 5). The complex two-step gelation mechanism of GG is confirmed by the presence of a double yielding behavior in the flow curve, which happens at low enough gellan gum concentration, only in the presence of divalent ions. This provides a rheological confirmation of the two-step aggregation process. In addition, we have proposed a reliable and highly reproducible preparation protocol for GG microgels. Starting from the pioneering work of Ref. [22] and the early results of Ref. [25], we nowput forward a rheological-based method that is able to systematically and consistently provide stable microgels that are very efficient in cleaning paper artefacts. We note that a recent work [58] reported the attempt to produce GG microgels by a different procedure, without success. Moreover our results have been provided not only GG but also on GGMA. Methacrylation is an interesting chemical modification, because enhances hydrophobicity of the polymer. In this way, GG and GGMA hydeogels or microgels are able to interact and adsorb not only hydrophilic by-products of cellulose degradation but also hydrophobic materials as synthetic adhesives of fats, representing an environmentally friendly strategy for preserving paper artworks, eliminating the need for potentially hazardous organic solvents in the removal of hydrophobic materials. The present work therefore provides a rheological overview of the gellan gum-based hydrogels and microgels and propose an optimal rheological window, in term of $G'$, that these systems have to satisfy to be efficient in cleaning. For hydrogels, $G'$ is slightly larger than for microgels, in order to make a rigid gel on the paper sheet that does not leave residues after an application under a weight that usually lasts one hour. Instead, for microgels, the employed solutions are much softer and act very quickly, so that they can still be successfully removed by the op- erators [25, 26]. Hence, we believe that the importance of our findings is manifold: i) the knowledge of the rheological properties of GG-based hydrogels/microgels plays a crucial role in choosing the best performant GG or GGMA hydrogel for paper artworks restoration ii) formulation of GG-based hydrogels/microgels with precisely controlled viscosity, enables to customize the application process to suit the specific requirements of the paper artifact while ensuring the preservation of paper fibers; iii) it fills the gap with existent studies on GG hydrogels and extend it to microgels iv) it envisages the potential benefits of these systems to a wide range of substrates as canvas and wood opening up new possibilities in the field; v) the knowledge of the viscoelastic properties could be crucial in designing materials with desired characteristics considering that a careful choice of the sample (polymer and salt concentration, type of salt) guarantees the best rheological performances in applications and broadens the potential benefits on different fields.

## 5. Acknowledgement

We acknowledge financial support from Regione Lazio through L.R. 13/08 Progetto Gruppo di Ricerca MICROARTE n. prot. A0375-2020-36515 and from ERC POC project MICROTECH (grant agreement no.101066434). RA acknowledges financial support under the National Recovery and Resilience Plan (NRRP), Mission 4, Component 2, Investment 1.1, Call for tender No. 104 published on 2.2.2022 by the Italian Ministry of University and Research (MUR), funded by the European Union – NextGenerationEU – Project PRIN 2022ZA77J2 ICARUS – CUP B53D23009010006.

## References

[1] G. Sworn, G. Sanderson, W. Gibson, Gellan gum fluid gels, Food Hydrocolloids 9 (4) (1995) 265–271.

[2] C. S. F. Picone, R. L. Cunha, Influence of ph on formation and properties of gellan gels, Carbohydrate Polymers 84 (1) (2011) 662–668.

[3] J. K. Oh, R. Drumright, D. J. Siegwart, K. Matyjaszewski, The development of microgels/nanogels for drug delivery applications, Progress in polymer science 33 (4) (2008) 448–477.

[4] E. R. Morris, K. Nishinari, M. Rinaudo, Gelation of gellan–a review, Food Hydrocolloids 28 (2) (2012) 373–411.

[5] S. Banerjee, S. Bhattacharya, Food gels: gelling process and new applications, Critical reviews in food science and nutrition 52 (4) (2012) 334–346.

[6] D. Saha, S. Bhattacharya, Characteristics of gellan gum based food gels, Journal of texture studies 41 (4) (2010) 459–471.

[7] P. Matricardi, C. Cencetti, R. Ria, F. Alhaique, T. Coviello, Preparation and characterization of novel gellan gum hydrogels suitable for modified drug release, Molecules 14 (9) (2009) 3376–3391.

[8] G. D'Arrigo, G. Navarro, C. Di Meo, P. Matricardi, V. Torchilin, Gellan gum nanohydrogel containing anti-inflammatory and anti-cancer drugs: a multi-drug delivery system for a combination therapy in cancer treatment, European Journal of Pharmaceutics and Biopharmaceutics 87 (1) (2014) 208–216.

[9] M. Milivojevic, I. Pajic-Lijakovic, B. Bugarski, A. K. Nayak, M. S. Hasnain, Gellan gum in drug delivery applications, Natural polysaccharides in drug delivery and biomedical applications (2019) 145–186.

[10] U. M. Musazzi, C. Cencetti, S. Franzé, N. Zoratto, C. Di Meo, P. Procacci, P. Matricardi, F. Cilurzo, Gellan nanohydrogels: novel nanodelivery systems for cutaneous administration of piroxicam, Molecular pharmaceutics 15 (3) (2018) 1028–1036.

[11] J. T. Oliveira, L. Martins, R. Picciochi, P. Malafaya, R. Sousa, N. Neves, J. Mano, R. Reis, Gellan gum: a new biomaterial for cartilage tissue engineering applications, Journal of Biomedical Materials Research Part A: An Official Journal of The Society for Biomaterials, The Japanese Society for Biomaterials, and The Australian Society for Biomaterials and the Korean Society for Biomaterials 93 (3) (2010) 852–863.

[12] A. M. Compaan, K. Song, Y. Huang, Gellan fluid gel as a versatile support bath material for fluid extrusion bioprinting, ACS applied materials & interfaces 11 (6) (2019) 5714–5726.

[13] S. Nagpal, S. K. Dubey, V. K. Rapalli, G. Singhvi, Pharmaceutical applications of gellan gum, in: Natural Polymers for Pharmaceutical Applications, Apple Academic Press, 2019, pp. 87–109.

[14] R. C. Sabadini, V. C. Martins, A. Pawlicka, Synthesis and characterization of gellan gum: chitosan biohydrogels for soil humidity control and fertilizer release, Cellulose 22 (2015) 2045–2054.

[15] M. C. García, M. C. Alfaro, N. Calero, J. Muñoz, Influence of gellan gum concentration on the dynamic viscoelasticity and transient flow of fluid gels, Biochemical engineering journal 55 (2) (2011) 73–81.

[16] J. Tang, M. Tung, Y. Zeng, Gelling properties of gellan solutions containing monovalent and divalent cations, Journal of Food Science 62 (4) (1997) 688–712.

[17] R. Chandrasekaran, R. P. Millane, S. Arnott, E. D. Atkins, The crystal structure of gellan, Carbohydrate Research 175 (1) (1988) 1–15.

[18] E. Miyoshi, T. Takaya, K. Nishinari, Rheological and thermal studies of gel-sol transition in gellan gum aqueous solutions, Carbohydrate Polymers 30 (2-3) (1996) 109–119.

[19] M. Diener, J. Adamcik, A. Sánchez-Ferrer, F. Jaedig, L. Schefer, R. Mezzenga, Primary, secondary, tertiary and quaternary structure levels in linear polysaccharides: from random coil, to single helix to supramolecular assembly, Biomacromolecules 20 (4) (2019) 1731–1739.

[20] M. Diener, J. Adamcik, J. Bergfreund, S. Catalini, P. Fischer, R. Mezzenga, Rigid, fibrillar quaternary structures induced by divalent ions in a carboxylated linear polysaccharide, ACS Macro Letters 9 (1) (2020) 115–121.

[21] X. Yang, M. Kimura, Q. Zhao, K. Ryo, F. B. A. Descallar, S. Matsukawa, Gelation of gellan induced by trivalent cations and coexisting trivalent with monovalent cations studied by rheological and dsc measurements, Carbohydrate Polymers (2024) 122485.

[22] M. Caggioni, P. T. Spicer, D. L. Blair, S. E. Lindberg, D. Weitz, Rheology and microrheology of a microstructured fluid: The gellan gum case, Journal of Rheology 51 (5) (2007) 851–865.

[23] M. Paulsson, H. Hägerström, K. Edsman, Rheological studies of the gelation of deacetylated gellan gum (gelrite®) in physiological conditions, European journal of pharmaceutical sciences 9 (1) (1999) 99–105.

[24] M. Nickerson, A. Paulson, R. Speers, Rheological properties of gellan solutions: effect of calcium ions and temperature on pre-gel formation, Food hydrocolloids 17 (5) (2003) 577–583.

[25] B. Di Napoli, S. Franco, L. Severini, M. Tumiati, E. Buratti, M. Titubante, V. Nigro, N. Gnan, L. Micheli, B. Ruzicka, et al., Gellan gum microgels as effective agents for a rapid cleaning of paper, ACS applied polymer materials 2 (7) (2020) 2791–2801.

[26] L. Severini, S. Franco, S. Sennato, E. Paialunga, L. Tavagnacco, L. Micheli, R. Angelini, E. Zaccarelli, C. Mazzuca, Methacrylated gellan gum microgels: the frontier of gel-based cleaning system, Submitted (2024).

[27] N. Khaksar-Baghan, A. Koochakzaei, Y. Hamzavi, An overview of gel-based cleaning approaches for art conservation, Heritage Science 18 (48) (2024).

[28] L. Micheli, C. Mazzuca, A. Palleschi, G. Palleschi, Development of a diagnostic and cleaning tool for paper artworks: a case of study, Microchemical Journal 126 (2016) 32–41.

[29] C. Mazzuca, L. Micheli, M. Carbone, F. Basoli, E. Cervelli, S. Iannuc-celli, S. Sotgiu, A. Palleschi, Gellan hydrogel as a powerful tool in paper cleaning process: A detailed study, Journal of colloid and interface science 416 (2014) 205–211.

[30] L. Severini, L. Tavagnacco, R. Angelini, S. Franco, M. Bertoldo, M. Calosi, L. Micheli, S. Sennato, E. Chiessi, B. Ruzicka, et al., Methacrylated gellan gum hydrogel: a smart tool to face complex problems in the cleaning of paper materials, Cellulose 30 (16) (2023) 10469–10485.

[31] H. Grasdalen, O. Smidsrød, Gelation of gellan gum, Carbohydrate polymers 7 (5) (1987) 371–393.

[32] G. Robinson, C. E. Manning, E. R. Morris, Conformation and physical properties of the bacterial polysaccharides gellan, welan, and rhamsan, in: Food polymers, gels and colloids, Elsevier, 1991, pp. 22–33.

[33] G. Robinson, C. E. Manning, E. R. Morris, Conformation and physical properties of the bacterial polysaccharides gellan, welan, and rhamsan, in: Food polymers, gels and colloids, Elsevier, 1991, pp. 22–33.

[34] L. Tavagnacco, E. Chiessi, L. Severini, S. Franco, E. Buratti, A. Capocefalo, F. Brasili, A. Mosca Conte, M. Missori, R. Angelini, et al., Molecular origin of the two-step mechanism of gellan aggregation, Science Advances 9 (10) (2023) eadg4392.

[35] A. Ahuja, A. Potanin, Y. M. Joshi, Two step yielding in soft materials, Advances in Colloid and Interface Science 282 (2020) 102179.

[36] K. Pham, G. Petekidis, D. Vlassopoulos, S. Egelhaaf, W. Poon, P. Pusey, Yielding behavior of repulsion-and attraction-dominated colloidal glasses, Journal of Rheology 52 (2) (2008) 649–676.

[37] N. Koumakis, G. Petekidis, Two step yielding in attractive colloids: transition from gels to attractive glasses, Soft Matter 7 (6) (2011) 2456–2470.

[38] H.-N. Xu, Y.-H. Li, Decoupling arrest origins in hydrogels of cellulose nanofibrils, ACS omega 3 (2) (2018) 1564–1571.

[39] C. Mazzuca, L. Micheli, R. Lettieri, E. Cervelli, T. Coviello, C. Cencetti, S. Sotgiu, S. Iannuccelli, G. Palleschi, A. Palleschi, How to tune a paper cleaning process by means of modified gellan hydrogels, Microchemical Journal 126 (2016) 359–367.

[40] S. Iannuccelli, S. Sotgiu, Wet treatments of works of art on paper with rigid gellan gels, The book and paper group annual 29 (2010) (2010) 25–39.

[41] M. Rubinstein, R. H. Colby, Polymer Physics, Oxford University Press, 2003.

[42] R. Kohlrausch, Thermoresponsive poly-(N-isopropylmethacrylamide) microgels: tailoring particle size by interfacial tension control, Pogg. Ann. Phys. Chem. 91 (1854) 179–214.

[43] D. C. W. G. Williams, Non-symmetrical dielectric relaxation behavior arising from a simple empirical decay function, J. Chem. Soc. Faraday Trans. 66 (1970) 80–85.

[44] F. Puoti, A. V. Jervis, R. Ciabattoni, E. Cossa, A. Di Giovanni, M. R. Giuliani, M. Iodele, Evaluation of Leather Cleaning with a Rigid Hidrogel of Gellan Gum on Two Composite Amharic Shields from the Museo Nazionale Preistorico Etnografico" Luigi Pigorini", Rome, Archetype Publications, 2017.

[45] M. C. García, M. C. Alfaro, J. Muñoz, Rheology of sheared gels based on low acyl-gellan gum, Food Science and Technology International 22 (4) (2016) 325–332.

[46] N. Calero, M. Alfaro, M. Lluch, M. Berjano, J. Munoz, Rheological behavior and structure of a commercial esterquat surfactant aqueous system, Chemical Engineering & Technology: Industrial Chemistry-Plant Equipment-Process Engineering-Biotechnology 33 (3) (2010) 481–488.

[47] A. H. Clark, Gels and gelling, Physical chemistry of foods 7 (1992) 263–305.

[48] A. Rodrıguez-Hernández, S. Durand, C. Garnier, A. Tecante, J. L. Doublier, Rheology-structure properties of gellan systems: evidence of network formation at low gellan concentrations, Food Hydrocolloids 17 (5) (2003) 621–628.

[49] A. P. Safronov, I. S. Tyukova, G. V. Kurlyandskaya, Coil-to-helix transition of gellan in dilute solutions is a two-step process, Food hydrocolloids 74 (2018) 108–114.

[50] S. J. Pérez-Campos, N. Chavarría-Hernández, A. Tecante, M. Ramírez-Gilly, A. I. Rodríguez-Hernández, Gelation and microstructure of dilute gellan solutions with calcium ions, Food hydrocolloids 28 (2) (2012) 291–300.

[51] L. Dai, X. Liu, Y. Liu, Z. Tong, Concentration dependence of critical exponents for gelation in gellan gum aqueous solutions upon cooling, European Polymer Journal 44 (12) (2008) 4012–4019.

[52] M. Tako, T. Teruya, Y. Tamaki, T. Konishi, Molecular origin for rheological characteristics of native gellan gum, Colloid and Polymer Science 287 (2009) 1445–1454.

[53] L. Dai, X. Liu, Z. Tong, Critical behavior at sol–gel transition in gellan gum aqueous solutions with kcl and cacl2 of different concentrations, Carbohydrate Polymers 81 (2) (2010) 207–212.

[54] H. H. Winter, F. Chambon, Analysis of linear viscoelasticity of a crosslinking polymer at the gel point, Journal of rheology 30 (2) (1986) 367–382.

[55] V. Y. Grinberg, T. V. Burova, N. V. Grinberg, A. Y. Mashkevich, I. G. Plashchina, A. I. Usov, N. P. Shusharina, A. R. Khokhlov, L. Navarini, A. Cesàro, Thermodynamics of the double helix-coil equilibrium in tetramethylammonium gellan: High-sensitivity differential scanning calorimetry data, Macromolecular Bioscience 3 (3-4) (2003) 169–178.

[56] A. Fernandez-Nieves, H. Wyss, J. Mattsson, D. A. Weitz, Microgel suspensions: fundamentals and applications, John Wiley & Sons, 2011.

[57] C. Corsaro, D. Mallamace, S. Vasi, L. Pietronero, F. Mallamace, M. Missori, The role of water in the degradation process of paper using 1 h hr-mas nmr spectroscopy, Physical Chemistry Chemical Physics 18 (48) (2016) 33335–33343.

[58] G. B. Messaoud, E. Stefanopoulou, M. Wachendörfer, S. Aveic, H. Fischer, W. Richtering, Structuring gelatin methacryloyl–dextran hydrogels and microgels under shear, Soft Matter 20 (4) (2024) 773–787.